\documentclass{appolb}
\usepackage{graphicx} 

\usepackage{cite} 
\usepackage{upgreek} 
\usepackage{amsmath} 

\newcommand{\electron}{\ensuremath{\mathrm{e}}}
\newcommand{\eplus}{\ensuremath{\mathrm{e}^{+}}}
\newcommand{\eminus}{\ensuremath{\mathrm{e}^{-}}}
\newcommand{\photon}{\ensuremath{\upgamma}}
\newcommand{\muon}{\ensuremath{\upmu}}
\newcommand{\muplus}{\ensuremath{\upmu^+}}
\newcommand{\muminus}{\ensuremath{\upmu^-}}
\newcommand{\piplus}{\ensuremath{\uppi^+}}
\newcommand{\piminus}{\ensuremath{\uppi^-}}
\newcommand{\pizero}{\ensuremath{\uppi^0}}
\newcommand{\U}{\ensuremath{\mathrm{U}}}
\newcommand{\Uboson}{U~boson}
\newcommand{\phimeson}{\ensuremath{\upphi}}
\newcommand{\etameson}{\ensuremath{\upeta}}

\newcommand{\eeUgUee}{\ensuremath{\eplus\eminus \to \mathrm{U}\upgamma}, \ensuremath{\mathrm{U} \to \eplus\eminus}}

\newcommand{\eV}{{e\kern-.07em V}}

\newcommand{\MeVc}{{\rm \,M\eV\kern-.09em /\kern-.08em c}}
\newcommand{\MeVcc}{{\rm \,M\eV\kern-.09em /\kern-.08em c$^{2}$}}

\newcommand{\GeVcc}{{\rm \,G\eV\kern-.09em /\kern-.08em c$^{2}$}}

\mathchardef\mhyphen="2D

\begin{document}
\title{
Search for the \Uboson{} in the process \\$\eplus\eminus \to \U\photon$, 
$\U \to \eplus\eminus$ with the KLOE detector
%
\let\thefootnote\relax\footnote{Submitted to \emph{Acta Physica Polonica B} on 24 October 2014.}
}
\author{Anthony Palladino
\address{Laboratori Nazionali di Frascati, Frascati, Italy}
\\
\address{on behalf of the KLOE-2 collaboration}
}
\maketitle
\begin{abstract}

Dark Matter and Dark Energy are two of the most fundamental open questions 
in physics today. 
The existence of a light dark-force mediator has been hypothesized 
as a possible explanation for several unexplained physical phenomena.
A new search for this mediator, the dark photon \U{}, is underway using 
data collected with the KLOE detector at DA$\Phi$NE. We describe the
strategy we will use in our search for a resonant peak in the electron-positron invariant
mass spectrum from the process $\eplus\eminus \to \U \photon$ 
with $\U \to \eplus\eminus$. So far we found no evidence for 
the process and set a preliminary upper limit on the level of mixing 
between the secluded dark sector and the standard model. 
\end{abstract}
\DOI{10.5506/APhysPolB.46.1001}

\section{Introduction}
A series of unexpected astrophysical observations have failed to find
explanations in terms of standard astrophysical or particle physics
models~\cite{integral,pamela,ams,atic,fermi,hess1,hess2,dama1,dama2,cogent}. 
Each of these anomalies can be explained, however, if there exists a dark
weakly interacting massive particle, WIMP, belonging to a secluded gauge 
sector~\cite{dieci,undici,dodici,tredici,diciannove}. A dark vector boson, \U{},
an abelian gauge field, may couple the secluded sector
to the Standard Model through its kinetic mixing with the Standard Model electroweak hypercharge gauge field,
$\mathcal{L}_{\mathrm{mix}} = -\,{}^{\varepsilon^2}\!\!/_{2} \, F^{\mathrm{EW}}_{ij}F^{ij}_{\mathrm{Dark}}$. The
kinetic mixing parameter, $\varepsilon$, is expected to be of the 
order $10^{-4}$--$10^{-2}$ which allows for  
observable effects in $\mathcal{O}$(GeV)-energy \eplus{}\eminus{}
colliders \cite{venti,ventuno,ventidue}. The
\Uboson{} might be produced in such collider experiments via several 
processes: $\mathrm{V} \to \mathrm{P} \U$ decays, where $\mathrm{V}$ and $\mathrm{P}$
are vector and pseudoscalar mesons, 
$\eplus\eminus \to \U\photon$ with $\U \to \ell^+\ell^-$, 
where $\ell = \mathrm{e}$ or $\upmu$, 
and $\eplus\eminus \to \U \mathrm{h}^{\prime}$ (dark Higgsstrahlung), where $\mathrm{h}^{\prime}$ 
is a Higgs-like particle responsible for breaking the hidden symmetry.

\section{The KLOE detector}
The KLOE experiment operated from 2000 to 2006 at DA$\Phi$NE, the Frascati 
$\upphi$ factory. DA$\Phi$NE is an \eplus{}\eminus{} collider 
running mainly at a center-of-mass energy of $\sim$1.0195~GeV, 
the mass of the $\upphi$ meson. 
Equal energy electron and positron beams
collide at an angle of $\sim$25~mrad, producing $\upphi$ mesons nearly 
at rest. The
detector consists of a large cylindrical Drift Chamber (DC)~\cite{kloe_drift_chamber}, 
providing a momentum resolution of
$\sigma_\bot/p_\bot \approx$ 0.4, surrounded by a
lead-scintillating fiber electromagnetic calorimeter (EMC)~\cite{kloe_calorimeter} providing
an energy resolution of $\sigma_E/E = 5.7\%/\sqrt{E (\mathrm{GeV})}$  
and a time resolution of $\sigma_t$ = 57~$\mathrm{ps}/\sqrt{E (\mathrm{GeV})} \, \oplus$~100~ps.
A superconducting coil around the EMC provides a 0.52~T field.

\section{\Uboson{} searches by KLOE-2}
The KLOE-2 Collaboration has completed three searches for a dark photon.
The first two searched for \Uboson{} in vector meson decays $\mathrm{V}\to\mathrm{P}\U$, where 
$\phimeson\to\etameson\U$, $\U\to\eplus\eminus$ with the pseudoscalar meson decaying
via $\etameson\to\piplus\piminus\pizero$~\cite{kloe_phi_eta_1} and 
 $\etameson\to\pizero\pizero\pizero$~\cite{kloe_phi_eta_2}.
KLOE-2 provided another limit for \Uboson{} production using the process
$\eplus\eminus\to\U\photon$, $\U\to\muplus\muminus$~\cite{kloe_mmg}.
A fourth dark force analysis has been performed by KLOE-2 by searching for the \Uboson{} in the dark 
Higgsstrahlung process, $\eplus\eminus \to \U \mathrm{h}^{\prime}$. A preliminary limit on 
the product of the dark coupling strength and the kinetic mixing strength,
$\alpha_{\mathrm{D}} \times \varepsilon^{2}$, will be published soon.

\section{U boson search in \eeUgUee}
The first three analyses produced excellent limits in the parameter space $\varepsilon^{2}$
versus $m_{\U}$, but some values of $\varepsilon$ and $m_{\U}$ that can explain the $(g\!-\!2)_{\muon}$ 
anomaly have not yet been excluded. In particular we would like to probe the range $15<m_{\U}<50$~\MeVcc{} to
either find evidence for an explanation of the muon anomaly or completely exclude the dark photon as
a possible explanation. 
At an $\eplus\eminus$ collider like DA$\Phi$NE, it's possible that the electron and positron can 
annihilate, or scatter, producing a U~boson and a photon, with the 
decay of the U~boson into a pair of leptons. 
Unlike the previous KLOE-2 limits, the sensitivity from the $\eeUgUee$ 
channel is expected to increase as $m_{\U}$ aproaches $2m_{\electron}$ due to the dramatic increase
in the \Uboson{} production cross section, 
\begin{equation}
\label{eq:ubos_cross_sec}
\sigma(\eplus\eminus \to \U \to \ell^{+}\ell^{-},s') = \frac{12\pi\Gamma(\mathrm{U}\to \eplus\eminus)\Gamma(\mathrm{U}\to \ell^{+}\ell^{-})}
                         {\left(s'-m_{\mathrm{U}}^{2}\right)^{2}+m_{\mathrm{U}}^{2}\Gamma_{\mathrm{Total}}^2}
\end{equation}
where we have electrons as our final-state leptons ($\ell = \electron$) and
$\Gamma_{\mathrm{Total}} = \Gamma( \mathrm{U} \to \eplus\eminus )
                          +\Gamma( \mathrm{U} \to \upmu^+\upmu^-)
                          +\Gamma( \mathrm{U} \to \mathrm{hadrons})$ is the total width.

A new KLOE-2 analysis is underway which proposes to search for \Uboson{} production in the 
process $\eplus\eminus\to\U\photon$, $\U\to\eplus\eminus$. 
The 3 final-state particles of this process are the same as radiative Bhabha scattering.
The distinct feature we are searching for is a Breit-Wigner resonant production peak
(at the U~boson mass) in the invariant-mass distribution of the $\eplus\eminus$ pair.
To search for a \Uboson{} produced at a fixed-energy \eplus\eminus{} collider
we use initial-state radiation (ISR) to reduce the center of mass
energy and thereby scan the range of possible \Uboson{} masses down to $2m_{\electron}$.
The process consists of finite-width effects for $s$-channel annihilation subprocesses,
non-resonant $t$-channel U~boson exchange, and $s$-$t$ interference contributions. 
The finite-width effects are order $\Gamma_{\mathrm{U}}/m_{\mathrm{U}}$ on the integrated
cross section so are much smaller than any potential resonance we would observe, but they
are critical from a phenomonological perspective and are properly taken into account in the 
Monte Carlo simulation~\cite{babayaga6}.
The non-resonant $t$-channel effects would not produce the Breit-Wigner peak in the 
invariant mass distribution but could, in principle, show up in analyses of angular 
distributions or asymmetries. The KLOE-2 analysis will focus exclusively on resonant $s$-channel 
\Uboson{} production.

Using about 1.5~fb$^{-1}$ of KLOE data collected during 2004--2005 we will search for
\Uboson{} production in a sample of radiative Bhabha scattering events.
The strategy is to select events with the final-state electron, positron, and photon, all emitted at
large angle ($55^{\circ}<\theta<125^{\circ}$) with respect to the beam axis,
such that they are explicitly detected in the barrel of
the calorimeter.   
The $m_{\mathrm{track}}$ variable, computed using energy and momentum conservation, 
with the assumption of equal-mass oppositely-charged particles, will be used to separate
electrons from the more massive muons and pions.

We will use Monte Carlo (MC) simulations to estimate the level of background contamination 
due to the following processes: $\eplus\eminus \to \muplus\muminus\photon$, 
$\eplus\eminus \to \piplus\piminus\photon$, $\eplus\eminus \to \photon\photon$ (where one photon 
converts into an $\eplus\eminus$ pair), 
and $\eplus\eminus \to \upphi \to \uprho\uppi^{0} \to \piplus\piminus\uppi^{0}$,
as well as other $\upphi$ decays.
Due to the KLOE detector's excellent efficiency at detecting electrons and distinguishing them from 
heavier charged particles, we estimate that the sum of all background processes 
is typically less than 1\% in the $m_{\mathrm{ee}}$ distribution.
None of the background shapes are peaked, 
eliminating the possiblility of a background mimicking the resonant \Uboson{} signal.

\begin{figure}[h]
\begin{minipage}[t]{0.5\linewidth}
\centering
\includegraphics[width=\linewidth]{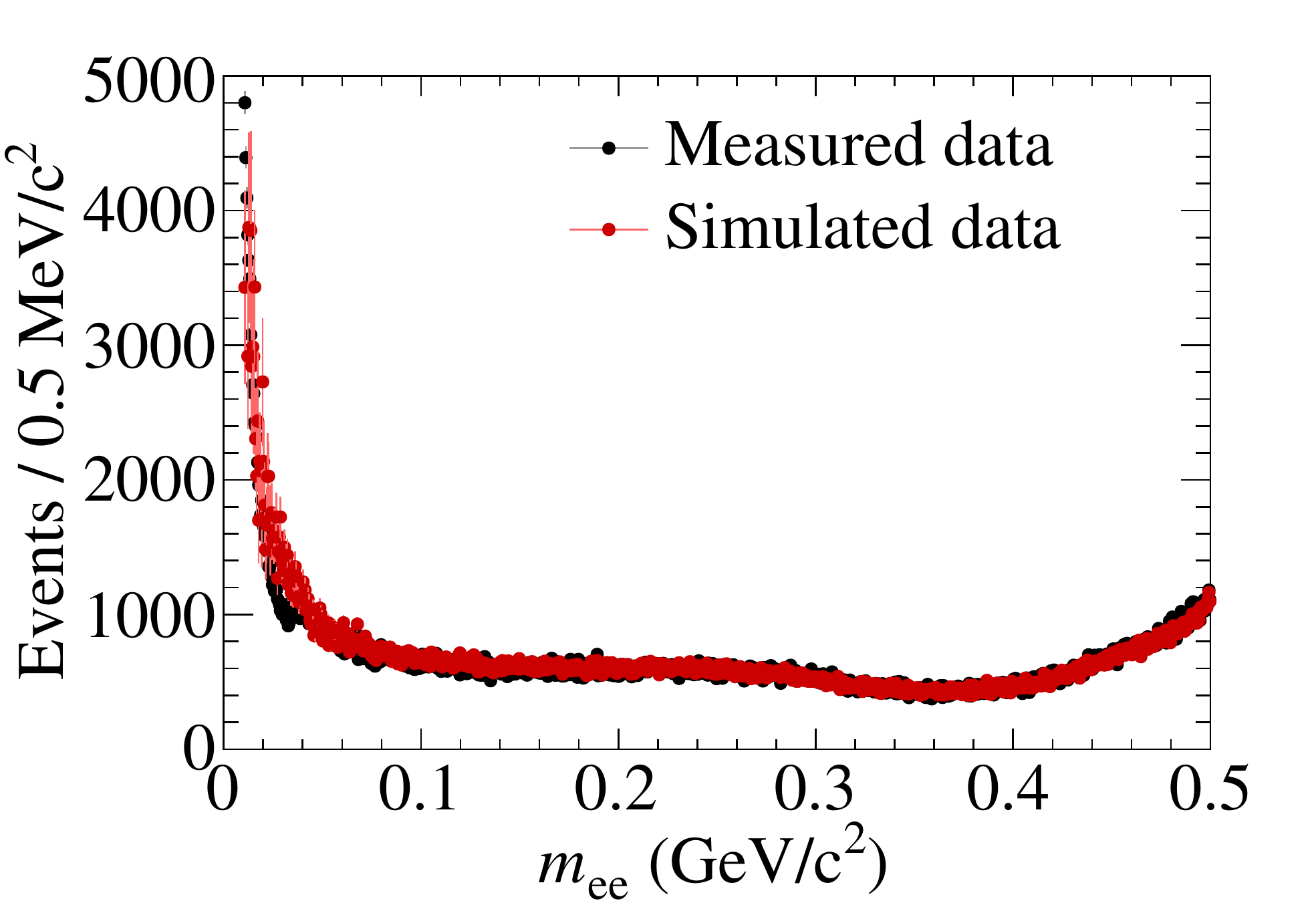}
\caption{Dielectron invariant mass distribution from 
   KLOE measurement data compared to our Monte Carlo simulation
   using the {\sc Babayaga-NLO} event generator.}
\label{Fig:m_ee_distribution}
\end{minipage}
\hspace{0.5cm}
\begin{minipage}[t]{0.5\linewidth}
\centering
\includegraphics[width=\linewidth]{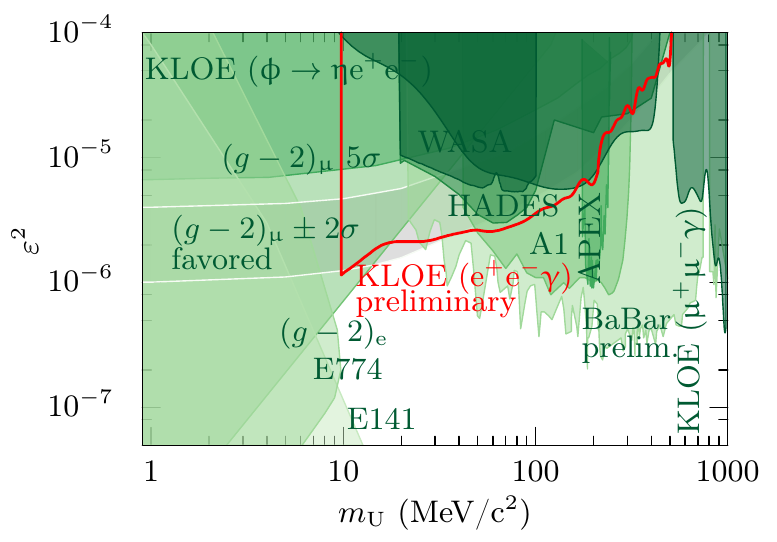}
\caption{Preliminary exclusion limit on the kinetic mixing parameter 
squared as a function of the \Uboson{} mass. This limit is not the final result. 
The gray band indicates the mixing levels and \Uboson{} masses that could explain the 
discrepancy observed between the measurement and SM calculation of the muon $(g\!-\!2)_{\upmu}$.}
\label{Fig:limits}
\end{minipage}
\end{figure}

Several Monte Carlo event generators for radiative Bhabha scattering fail to accurately reproduce the 
physics at the dielectron mass threshold due to numerical instabilities in  
integrations of the form $\frac{1}{q^2}\sqrt{1 - \frac{4m^2}{q^2}}$. 
Due to the three order-of-magnitude difference
between the electron mass and the center-of-mass energy of the collision, numerical instabilities
arise as $q^{2}$ approaches threshold where the square root gives 0, but as $q^{2}$ becomes larger than $4m^{2}$ 
the factor $1/q^{2}$ becomes dominant.
These problems are apparent when the simulated cross section fails to show the significant rise at threshold.
Together with the authors of {\sc Babayaga} we modified 
the {\sc Babayaga-NLO}~\cite{babayaga1,babayaga2,babayaga3,babayaga4,babayaga5,babayaga6}
event generator and implemented it into our full KLOE simulation such that the weighted events are 
distributed throughout the phase space with the square of the matrix element providing the correct weight.
The good agreement between our MC simulation using the new event generator and our selected data sample
is shown in Fig.~\ref{Fig:m_ee_distribution}.

No signal peak has been observed so far.
A preliminary excercise was performed on measured data 
using the CLS technique~\cite{cls} to determine a preliminary limit on the number of signal
\Uboson{} events, $N_{\U}$, at 90\% confidence level. 
Chebyshev polynomials were fit to the measured data ($\pm$15$\sigma$), 
excluding the signal region of interest ($\pm$3$\sigma$), and were used as the background. 
A Breit-Wigner peak smeared with the invariant mass resolution was
used as the signal.

We then translated this limit on $N_{\U}$ to a 90\% confidence level limit on the kinetic mixing parameter 
as a function of $m_{\mathrm{ee}}$ as~\cite{m_mumu_article}
\begin{equation}
\label{eq:epsilon2}
\varepsilon^2\!\left(m_{\mathrm{ee}}\right) = \frac{ N_{\U}\!\left(m_{\mathrm{ee}}\right) }{\epsilon_{\mathrm{eff}}\!\left(m_{\mathrm{ee}}\right) }
                                   \frac{1}{ H\!\left(m_{\mathrm{ee}}\right) \, I\!\left(m_{\mathrm{ee}}\right) \, L } \,\,\, ,
\end{equation}
where the radiator function $H\!\left(m_{\mathrm{ee}}\right)$ was extracted from 
$\mathrm{d}\sigma_{\mathrm{ee}\photon}/\mathrm{d}m_{\mathrm{ee}} = H\!\left(m_{\mathrm{ee}},s,\mathrm{cos}(\theta_\photon)\right) \cdot \sigma^{\mathrm{QED}}_{\mathrm{ee}}\!\left(m_{\mathrm{ee}}\right)$
using the {\sc Phokhara} MC simulation~\cite{PHOKHARA} to determine the radiative
differential cross section, $I\!\left(m_{\mathrm{ee}}\right)$ is the integral of 
the cross section (\ref{eq:ubos_cross_sec}),
and $L \simeq 1.5$~fb$^{-1}$ is the integrated luminosity. The selection efficiency, $\epsilon_{\mathrm{eff}}$, was obtained
from a {\sc Babayaga} MC simulation where the radiative Bhabha scattering was only allowed
to proceed via the annihilation channel, since that is the channel in which the \Uboson{}
Breit-Wigner resonance would occur; the $t$-channel ultimately becoming a background.
Our preliminary limit is shown in Fig.~\ref{Fig:limits} along with the limit 
from $\left( g-2 \right)_\upmu$ at 5$\sigma$,
E141~\cite{e141_1987},
E774~\cite{e774_1991}, 
KLOE($\upphi \to \eta \U$, $\U \to \eplus\eminus$)~\cite{kloe_phi_eta_1,kloe_phi_eta_2}, 
Apex~\cite{Apex2011},        
WASA~\cite{wasa2013},
HADES~\cite{hades2013},
A1~\cite{A1_2014},
KLOE($\eplus\eminus \to \U\photon$, $\U \to \muplus\muminus$)~\cite{m_mumu_article},
and a preliminary result from BaBar~\cite{babar2014}.

\section{Conclusions}
We outlined our strategy for a new dark gauge \Uboson{} search in the process $\eplus\eminus \to \U\photon$ 
with $\U \to \eplus\eminus$ using $\sim$1.5~fb$^{-1}$ of KLOE data collected in 2004--2005.
After a preliminary excercise, we found no evidence for the existence of a \Uboson{} and set 
a preliminary upper limit at $10^{-5}$--$10^{-7}$ on the level of kinetic mixing with the
Standard Model as a function of the \Uboson{} mass in the range 10--520~MeV/c$^2$.
A final result is forthcoming and should extend the limit closer to the dielectron mass threshold.
The upgraded KLOE-2 experiment~\cite{KLOE2}, currently running, uses a new cylindrical
GEM inner tracker~\cite{balla} providing higher resolution interaction vertexing, and plans to collect 
upwards of 10~fb$^{-1}$ of data. The increased statistical power and tracking/vertexing sensitivity 
will allow KLOE-2 to significantly extend our present limits.

\section{Acknowledgements}
We warmly thank our former KLOE colleagues for the access to the data collected during the KLOE data taking campaign.
We thank the DA$\Phi$NE team for their efforts in maintaining low background running conditions and their collaboration during all data taking. We want to thank our technical staff: 
G.F.~Fortugno and F.~Sborzacchi for their dedication in ensuring efficient operation of the KLOE computing facilities; 
M.~Anelli for his continuous attention to the gas system and detector safety; 
A.~Balla, M.~Gatta, G.~Corradi and G.~Papalino for electronics maintenance; 
M.~Santoni, G.~Paoluzzi and R.~Rosellini for general detector support; 
C.~Piscitelli for his help during major maintenance periods. 
This work was supported in part by the EU Integrated Infrastructure Initiative Hadron Physics Project under contract number RII3-CT- 2004-506078; by the European Commission under the 7$^{\mathrm{th}}$ Framework Programme through the `Research Infrastructures' action of the `Capacities' Programme, Call: FP7-INFRASTRUCTURES-2008-1, Grant Agreement No. 227431; by the Polish National Science Centre through the Grants No. 
DEC-2011/03/N/ST2/02641, 
2011/01/D/ST2/00748,
2011/03/N/ST2/02652,
2013/08/M/ST2/00323,
and by the Foundation for Polish Science through the MPD programme and the project HOMING PLUS BIS/2011-4/3.




\vspace{0.3cm}
\begin{thebibliography}{0}
\bibitem{integral}
P.~Jean, et al., 
Astron. Astrophys., {\bf 407} (2003), L55

\bibitem{pamela}
O. Adriani, et al.,
Nature, {\bf 458} (2009), 607

\bibitem{ams}
M. Aguilar, et al.,
Phys. Rev. Lett. {\bf 110} (2013), 141102

\bibitem{atic}
J. Chang, et al.,
Nature, {\bf 456} (2008), 362

\bibitem{fermi}
A. A. Abdo, et al.,
Phys. Rev. Lett., {\bf 102} (2009), 181101

\bibitem{hess1}
F. Aharonian, et al.,
Phys. Rev. Lett., {\bf 101} (2008), 261104

\bibitem{hess2}
F. Aharonian, et al.,
Astron. Astrophys., {\bf 508} (2009), 561

\bibitem{dama1}
R. Bernabei, et al.,
Int. J. Mod. Phys. D, {\bf 13} (2004), 2127

\bibitem{dama2}
R. Bernabei, et al.,
Eur. Phys. J. C, {\bf 56} (2008), 333

\bibitem{cogent}
C. E. Aalseth, et al.,
Phys. Rev. Lett., {\bf 107} (2011), 141301

\bibitem{dieci}
M. Pospelov, A. Ritz, M. B. Voloshin,
Phys. Lett. B, {\bf 662} (2008), 53

\bibitem{undici} 
N. Arkani-Hamed, et al.,
Phys. Rev. D, {\bf 79} (2009), 015014

\bibitem{dodici} 
D. S. M. Alves, et al.,
Phys. Lett. B, {\bf 692} (2010), 323

\bibitem{tredici}
M. Pospelov, A. Ritz,
Phys. Lett. B, {\bf 671} (2009), 391

\bibitem{diciannove}
N. Arkani-Hamed, N. Weiner,
JHEP, {\bf 0812} (2008), 104

\bibitem{kloe_phi_eta_1}
F.~Achilli, et al. (KLOE-2 Collab.),
Phys. Lett. B, {\bf 706} (2012), 251--255

\bibitem{kloe_phi_eta_2}
D.~Babusci, et al. (KLOE-2 Collab.),
Phys. Lett. B, {\bf 720} (2013), 111--115

\bibitem{venti}
R. Essig, P. Schuster, N. Toro,
Phys. Rev. D, {\bf 80} (2009), 015003

\bibitem{ventuno}
B. Batell, M. Pospelov, A. Ritz,
Phys. Rev. D, {\bf 79} (2009), 115008

\bibitem{ventidue}
M. Reece, L.T. Wang,
JHEP, {\bf 0907} (2009), 051

\bibitem{kloe_drift_chamber}
M.~Adinolfi, et al.
Nucl. Instr. Meth. A, {\bf 488} (2002), 51

\bibitem{kloe_calorimeter}
M.~Adinolfi, et al.
Nucl. Instr. Meth. A, {\bf 482} (2002), 364


\bibitem{kloe_phietaU_2012}
KLOE-2 Collaboration, Phys. Lett. B, {\bf 706}, (2012) 251--255

\bibitem{kloe_phietaU_2013}
KLOE-2 Collaboration, Phys. Lett. B, {\bf 720}, (2013) 111--115

\bibitem{kloe_mmg}
KLOE-2 Collaboration, Phys. Lett. B, {\bf 736}, (2014) 459--464

\bibitem{babayaga1}
G.~Balossini, et al., Nucl. Phys. B, {\bf 758}, (2006) 227--253

\bibitem{babayaga2}
G.~Balossini, et al., Phys. Lett. B, {\bf 663}, (2008) 209--313

\bibitem{babayaga3}
C.~M.~Carloni~Calame, et al., Nucl. Phys. Proc. Suppl. 131, (2004) 48--55

\bibitem{babayaga4}
C.~M.~Carloni~Calame, Phys. Lett. B, {\bf 520}, (2001) 16--24

\bibitem{babayaga5}
C.~M.~Carloni~Calame, et al., Nucl. Phys. B, {\bf 584}, (2000) 459--479

\bibitem{babayaga6}
L.~Barz\`{e}, et al., Eur. Phys. J. C, {\bf 71} (2011), 1680


\bibitem{cls} 
G.~C.~Feldman, R.~D.~Cousins, 
Physical Rev. D {\bf 57}, 3873 (1998)

\bibitem{m_mumu_article}
D.~Babusci, et al. (KLOE-2 Collaboration),
Phys. Lett. B {\bf 736} (2014), 459--464

\bibitem{PHOKHARA} 
H. Czy{\.z}, et al., Eur. Phys. J. C, {\bf 39} 411 (2005)

\bibitem{e141_1987}
E.~M.~Riordan, et al. (E141 Collaboration), 
Phys. Rev. Lett. {\bf 59} (1987), 755

\bibitem{e774_1991}
A.~Bross, et al. (E774 Collaboration), 
Phys. Rev. Lett. {\bf 67} (1991), 2942

\bibitem{Apex2011} 
S. Abrahamyan, et al. (APEX Collab.),
Phys. Rev. Lett. {\bf 107} (2011), 191804 

\bibitem{wasa2013}
P.~Adlarson, et al.  (WASA-at-COSY Collab.), 
Phys. Lett. B {\bf 726} (2013), 187

\bibitem{hades2013}
G.~Agakishiev, et al.  (HADES Collab.) 
Phys. Lett. B {\bf 731}, (2014), 265--271


\bibitem{A1_2014}
H.~Merkel, et al.  (A1 Collaboration),
Phys. Rev. Lett. 112 (2014), 221802



\bibitem{babar2014}
J.~P.~Lees, et al. (BaBar Collaboration), 
[arXiv:1406.2980] (2014)

\bibitem{KLOE2}
G.~Amelino-Camelia, et al.,
Eur. Phys. J. C, {\bf 68}, 619 (2010)

\bibitem{balla}
A.~Balla {\it et al.}, {\it JINST} {\bf 9}, C01014 (2014).


\end{thebibliography}
\end{document}